\documentclass[preprint,aps]{revtex4}
\input boxedeps.tex
\SetRokickiEPSFSpecial  %% dvips by Tom Rokicki
\HideDisplacementBoxes

\newcommand{\g}{\gamma}

\newcommand{\da}{\dagger}  % symbol for Hermitian conjugate dagger
\newcommand{\be}{\begin{equation}}
\newcommand{\eq}{\end{equation}}
\newcommand{\Tr}{{\rm \, Tr \!}}    %symbol for trace
\begin{document}
\mbox{SWAT/355}\hfil\\
%\draft
\title{Transverse lattice calculation of  
the pion light-cone wavefunctions}

\author{Simon Dalley}
\affiliation{Department of Physics, University of Wales Swansea \\
Singleton Park, Swansea SA2 8PP, United Kingdom}

\author{Brett van de Sande}
\affiliation{Geneva College, 3200 College Ave., Beaver Falls, PA~~15010}
%%\date{\today}

\begin{abstract}
We calculate the light-cone wavefunctions of the pion by 
solving the meson boundstate problem
in a coarse transverse lattice gauge theory using DLCQ. 
A large-$N_c$ approximation is made and the light-cone
Hamiltonian expanded in
massive dynamical fields at fixed lattice spacing. 
In contrast to earlier calculations, we include contributions
from states containing
many gluonic link-fields between the quarks.
The Hamiltonian is renormalised
by a combination of covariance conditions on boundstates 
and fitting the physical masses ${\cal M}_{\rho}$ and ${\cal M}_{\pi}$,
decay constant $f_{\pi}$, and the string tension $\sqrt{\sigma}$. 
Good covariance is obtained for the lightest $0^{-+}$ state, which we
identify with the pion. 
Many observables can be deduced  from its light-cone wavefunctions.
After perturbative evolution,
the quark valence structure function is found to be consistent with
the experimental structure function deduced from
Drell-Yan pi-nucleon data in the valence region $x > 0.5$. 
In addition, the pion distribution
amplitude is consistent with the experimental distribution deduced 
from the $\pi \gamma^* \gamma$
transition form factor and diffractive dissociation. 
A new observable we calculate is the probability for quark helicity 
correlation. We find a 45\% probability that the valence-quark helicities 
are aligned
in the pion.

\end{abstract}

\pacs{Valid PACS appear here.
{\tt$\backslash$\string pacs\{\}} should always be input,
even if empty.}
\maketitle

%%%%%%%%%%%%%%%%%%%%%%%%%%%%%%%%%%%%%%%%%%%%%%%%%%%%%%%%%%%%%%%%%%%%%%
%%%%%%%%%%%%%%%%%%%%%%%%%%%%%%%%%%%%%%%%%%%%%%%%%%%%%%%%%%%%%%%%%%%%%%
%%%%%%%%%%%%%%%%%%%%%%%%%%%%%%%%%%%%%%%%%%%%%%%%%%%%%%%%%%%%%%%%%%%%%%
\baselineskip .2in

\section{Introduction}
\label{intro}

Light-cone wavefunctions  encode all of the bound state
quark and gluonic properties of hadrons, including their momentum,
spin and flavour correlations, in the form of universal process- and 
frame-independent amplitudes (see, for example, Ref.~\cite{stan}). 
Hadronic observables represented as 
matrix elements of currents are easily expressed in terms of overlaps of
light-cone wavefunctions. To compute the wavefunctions, one must
diagonalise the light-cone Hamiltonian of QCD in a Fock space of quark
and gluonic degrees of freedom. A promising method to achieve this is
the transverse lattice formulation of gauge theory \cite{bard,rev}. In this
approach, the physical gluonic degrees of freedom are represented by
gauge-covariant links of colour flux on a lattice transverse to the
null-plane of quantisation. 
In this paper, we set up the method and solve for the  light-cone
wavefunctions of light mesons using a physically realistic truncation
of Fock space on a coarse lattice, spacing $\sim 2/3$ fm.
We obtain good covariance for the light-cone wavefunction of the lightest
meson, which we identify with the pion.
Results for the pion distribution amplitude (valence quark
wavefunction at small transverse separation) and distribution function
(valence quark probability at any transverse separation)
are consistent with the most recent experimental results in the valence
region of light-cone momenta. We find the distribution amplitude to be
\be
\phi_{\pi}(x)  =  
6 x (1-x) \left\{ 1 + 0.15(2)\, C_{2}^{3/2}(1-2x) + 0.04(1)\, C_{4}^{3/2}(1-2x)
\right\} 
 \ ,
\eq
while the distribution function is
\be
V_{\pi}(x)=
{(1-x)^{0.33(2)} \over x^{0.7(1)}}\left\{0.33(3)-1.1(2)\sqrt{x} + 2.0(3)x
  \right\} \; .
\eq
where $x$ is the quark light-cone momentum fraction carried in the pion.
The transverse renormalisation scale should be $0.5$~GeV if the first
moment of $V_{\pi}$ is to agree with experiment.  As a further application
of the light-cone wavefunctions, we also compute the probability for
a valence 
quark of momentum fraction $x$ to have its helicity correlated with that
of the anti-quark in the pion. 
We find a surprisingly large probability $\sim 45$~\%
for the quark and anti-quark helicities to be aligned, even though the pion is
spin 0.
These represent our main results.

Attempts to solve transverse lattice QCD
have been renewed in recent years for both the pure gauge theory 
\cite{grif,dv,mat1}
and mesons \cite{mat2,sd,mat3}. The most succesful approaches have
employed the original idea \cite{bard} of a
 $1/N_c$ and colour-dielectric expansion 
in dynamical fields to approximate the light-cone QCD Hamiltonian
on a coarse transverse lattice.
For pure gauge theory, to lowest non-trivial order of the expansion, 
requirements of vacuum stability, Lorentz and gauge invariance alone
were found to constrain
the coarse lattice Hamiltonian sufficiently accurately for first-principles
predictions of the glueball states \cite{dv}. Extension of this work to
light mesons  introduced quarks and imposed a (Tamm-Dancoff) restriction
on the number of link fields in Fock space \cite{mat2}. 
In previous calculations \cite{sd,mat3}, 
no more than one link field was allowed
in a meson. 
This effectively restricts the transverse size to
$<2/3$ fm,
which is unrealistic for light mesons. In this case, the 
correct Hamtiltonian could not be accurately identified using
Lorentz and gauge invariance alone.
Some phenomenology was also needed.

In this paper, we again use the lowest 
non-trivial order of the colour-dielectric
expansion of the hamiltonian, but relax the Tamm-Dancoff cut-off 
on the space of
states. This allows light mesons to
expand 
to their physical transverse size. It also means that 
one begins to take account of the full pure-glue dynamics in the meson sector.
While the results are now realistic, we find
that it is still necessary to use some phenomenological fitting
of masses and decay constants, in addition to optimizing Lorentz covariance, 
to obtain
unambiguous couplings in the coarse-lattice  Hamiltonian.
We believe this is due to the absence, in the currently employed 
transverse lattice Hamiltonian, of operators needed to optimize 
chiral symmetry. We show that such operators would occur at higher order
of the colour-dielectric expansion.
In the next section, we review and extend the previous work. 
Section~\ref{three} describes the procedure we employ for fixing 
the various couplings that appear in the Hamiltonian.
Finally, our results for pion observables are discussed in Section~\ref{four}.
Chiral symmetry issues are discussed in an Appendix.

\section{Transverse lattice mesons}
\label{two}
\subsection{Hamiltonian}

We introduce continuum light-cone co-ordinates $x^{\pm}= (x^0 \pm
 x^3)/\sqrt{2}$ and discretize the transverse coordinates 
${\bf  x}=(x^1,x^2)$ on a square lattice of spacing $a$. 
Lorentz indices $\mu, \nu $ are
split into light-cone indices $\alpha,\beta \in \{+,-\}$
and transverse indices $r,s\in \{1,2\}$. Subsequent analysis is done
to leading order of  the $1/N_c$ expansion of the gauge group $SU(N_c)$.
Quark fields $\Psi(x^+,x^-,{\bf x})$ in the fundamental
representation and  gauge potentials
 $A_{\alpha}(x^+,x^-,{\bf x})$ in the adjoint representation of
$SU(N_c)$ are associated to the sites of the
 transverse
lattice. Link fields $M_r(x^+,x^-,{\bf x})$, which we choose to be complex
 $N_c$x$N_c$ matrices,  are associated
with the directed link from ${\bf x}$ to ${\bf x} + a \hat{\bf
 r}$. They carry flux from site to site.
This use of disordered link variables implies that a coarse
transverse lattice is being considered. 

For finite spacing $a$, the Lagrangian can contain any
operators
that are local,  invariant under transverse lattice gauge symmetries
and under Poincar\'e symmetries manifestly
preserved by the lattice cut-off, and renormalisable by dimensional
counting with respect to the continuum co-ordinates $x^{\alpha}$. 
The objective is to obtain an approximation to the
 light-cone Hamiltonian operator $P^-$,
that may be diagonalised in a Fock state basis of the above fields.
This may be achieved by first fixing to the light-cone gauge $A_{-} =
0$, eliminating non-dynamical fields, then expanding the resulting
Hamiltonian in powers of the remaining dynamical fields. 
Truncation of
such a `colour-dielectric' expansion is a valid approximation provided
wavefunctions of interest (typically those of the lightest
eigenstates) are dominated by few-body Fock states. This is achieved
by working in a region of coupling space with  sufficiently heavy
dynamical fields. This in turn will be found to 
constrain the transverse lattice
spacing $a$ to be coarse  when 
symmetries and phenomenology are optimized.

The Lagrangian density we consider contains terms up to order  $M^4$ and 
$\overline{\Psi} M \Psi$ for the large-$N_c$ theory,
\begin{eqnarray}
L & = &  \sum_{{\bf x}} \int dx^- \sum_{\alpha, \beta = +,-}
\sum_{r=1,2} 
-{1 \over 2 G^2} \Tr \left\{ F^{\alpha \beta}({\bf x}) 
F_{\alpha \beta}({\bf x}) \right\}
 \nonumber
\\
&& + \Tr\left\{\overline{D}_{\alpha}M_r({\bf x}) (\overline{D}^{\alpha}
M_r({\bf x}))^{\dagger}\right\} 
\nonumber \\
&& - \mu_{b}^2  \Tr\left\{M_r M_r^{\da}\right\}
 + {\rm i} \overline{\Psi} 
\g^{\alpha} (\partial_{\alpha} + {\rm i} A_{\alpha}) \Psi - \mu_f
\overline{\Psi}\Psi 
\nonumber\\
&& +  {\rm i} \kappa_a \left( \overline{\Psi}({\bf x}) \g^{r} M_r({\bf x})
 \Psi({\bf x} + a \hat{\bf r}) 
- \overline{\Psi}({\bf x}) \g^{r} M_{r}^{\da}({\bf x}- a \hat{\bf r}) 
\Psi({\bf x} - a \hat{\bf r})
\right)\nonumber\\
&&
+ \kappa_s \left( \overline{\Psi}({\bf x}) M_r({\bf x})
 \Psi({\bf x} + a \hat{\bf r})+\overline{\Psi}({\bf x}) M_{r}^{\da}({\bf
x}- 
a \hat{\bf r})
 \Psi({\bf x} - a \hat{\bf r})\right) - V_{\bf x} \ , 
\label{ferlag}
\end{eqnarray}
where $F^{\alpha \beta}({\bf x})$ is the continuum field strength in the
$(x^0,x^3)$ planes at each ${\bf x}$,
\be
\overline{D}_{\alpha}M_r({\bf x}) = (\partial_{\alpha} + 
{\rm i} A_{\alpha} ({\bf x}))
        M_r({\bf x})-  {\rm i} M_r({\bf x})   A_{\alpha}({{\bf x}+a
 \hat{\bf r}})
\eq
and the link-field potential is 
\begin{eqnarray}
 V_{\bf x} & = & 
- {\beta\over N_c a^2} \sum_{r \neq s} 
  \Tr\left\{ M_{r} ({\bf x}) 
M_{s} ({\bf x} + a \hat{\bf r})
M_{r}^{\da}
({\bf x} + a \hat{\bf s} )  
M_{s}^{\da}({\bf x})
\right\} 
\nonumber \\
&& + {\lambda_1 \over a^{2} N_c} \sum_r  
\Tr\left\{ M_r M_r^{\da}
M_r M_r^{\da} \right\} 
+  {\lambda_2 \over a^{2} N_c}\sum_r  
\Tr\left\{ M_r ({\bf x}) M_r({\bf x} + a \hat{\bf r} )
M_r^{\da}({\bf x} + a \hat{\bf r} ) M_r^{\da} ({\bf x})\right\} \nonumber \\
&& +  {\lambda_4 \over a^{2} N_c}  
\sum_{\sigma=\pm 2, \sigma^\prime = \pm 1}
        \Tr\left\{ 
M_\sigma^{\da} M_\sigma M_{\sigma^\prime}^{\da} M_{\sigma^\prime} \right\} 
 \; . \label{pot1}
\end{eqnarray}
We have defined $M_{r} = M_{-r}^{\dagger}$ and hold
$\overline{G} \to {G} \sqrt{N_c}$ finite as $N_c \to \infty$.
To this action we could in principle add allowed operators at order
$M^6$, $(\overline{\Psi}\Psi)^2$, $\Psi M^2 \Psi$, and so on. It
should therefore be understood as the truncation of an expansion in
powers of the fields.  
Strictly speaking, this expansion should be performed for
the light-cone gauge-fixed Hamiltonian in terms of dynamical fields only. 

In the chiral representation, 
$\Psi^{\dagger} = (u_{+}^{*}, v_{+}^{*}, v_{-}^{*},
u_{-}^{*})/2^{1/4}$ decomposes
into 
complex fermion fields $v$ ($u$) with a helicity
subscript $h = \pm$ denoting the sign of the eigenvalue of
$\gamma^5$.
In light-cone gauge $A_- = 0$,
$A_{+}$ and $v_{\pm}$ are non-dynamical (independent of light-cone
time $x^+$) and 
are eliminated at the classical level using the equations of motion
\begin{eqnarray}
(\partial_{-})^2 A_{+} &  = & {G^2 \over 2} \left( J^+ - {1 \over N} \Tr
\ J^+ \right)\\
{\rm i}
\partial_{-} v_{h} & = & {{\mu_f} \over \sqrt{2} } F_{-h} \ \label{cons},
\end{eqnarray}
where we have defined
\begin{eqnarray}
F_{h}({\bf x}) & = & - u_{h}({\bf x}) + {\kappa_s \over \mu_f}
\sum_{r} \left( M_r({\bf x}) u_{h}({\bf x}+ a \hat{\bf r}) + 
M_{r}^{\da}({\bf x}-a \hat{\bf r} )u_{h}({\bf x}- a \hat{\bf r})
  \right) \nonumber \\ 
&&+  
{h {\rm i} \kappa_a \over  \mu_f} \left\{ M_{1}({\bf x} ) u_{-h}({\bf
x}+ a \hat{\bf 1})
- h{\rm i}M_{2}({\bf x} )) u_{-h}({\bf x}+ a \hat{\bf 2}) \right. \nonumber \\
&&   \left. - M_{1}^{\da}({\bf x}- a \hat{\bf 1} ) u_{-h}({\bf x}- a 
\hat{\bf 1})
+ h{\rm i}M_{2}^{\da}({\bf x}- a \hat{\bf 2} ) u_{-h}({\bf x}- a \hat{\bf 2})
\right\} \label{eff} , \\
J^{+}({\bf x}) &=& {\rm  i} \sum_{r}
\left(
M_r ({\bf x}) \stackrel{\leftrightarrow}{\partial}_{-} 
M_r^{\da}({\bf x})  + M_r^{\da}({\bf x} - a\hat{\bf r}) 
\stackrel{\leftrightarrow}{\partial}_{-} M_r({\bf x} - a\hat{\bf r})
\right)  \nonumber \\
&& + \sum_{h} u_{h}({\bf x})u_{h}^{\dagger}({\bf x}) \ , \label{jai}
\end{eqnarray}
The lightcone Hamiltonian, expressed in terms of the remaining 
dynamical fields $u_{\pm}({\bf x})$ and $M_{r}({\bf x})$, 
may be obtained from the action (\ref{ferlag}) 
in the standard way \cite{rev}  
\begin{eqnarray}
 P^-  &  = &  \int dx^- \sum_{{\bf x}} 
   {G^2 \over 4} \left(\Tr\left\{ 
              J^{+} \frac{1}{({\rm i} \partial_{-})^{2}} J^{+} \right\}
            -{1 \over N_{c}} 
        \Tr\left\{ J^+  \right\} {1 \over ({\rm i}\partial_{-})^{2} }
     \Tr\left\{ J^+ \right\} \right)  \nonumber \\
&& + {\mu_{f}^{2} \over 2} \sum_{h} 
\left( F_{h}^{\da} {1 \over {\rm i} \partial_-}
F_h \right) 
 +    V_{\bf x}[M]
\ . \label{ham} 
\end{eqnarray}

Under certain reasonable assumptions \cite{sd}, 
the Hamiltonian (\ref{ham}) is a truncation of the most general  
Hamiltonian to order $M^4$ and $u M u$. It also contains some, but not
all, allowed operators at order $u M^2 u$ and $ u^4 $. In particular,
it contains the combination $J^{+} \partial_{-}^{-2} J^{+}$, which 
is responsible for confinement in the lattice theory of states
singlet under residual $x^-$-independent gauge transformation \cite{bard}.  
The various parameters
$G,\mu_f,\kappa_a,\kappa_s,\mu_b,\lambda_1,\lambda_2,\lambda_4,
\beta$, 
as well as ones that would appear
at higher orders of the colour-dielectric expansion, 
are coupling constants that
will run with the cut-off(s) in the theory. 
In principle, this running could  be determined by performing
renormalisation group transformations 
from QCD at short distance scales.
However, on a coarse lattice, weak-coupling perturbation theory
is not available, and such an approach become unworkable.
One may also
treat the problem as an effective field theory, fixing
couplings phenomenologically.
Even in this case, one may constrain
the parameters from first principles by empirically tuning them to 
minimize the violation of continuum symmetries.  In the case of pure
gauge theories, at lowest order of the colour-dielectric expansion, this
gave a quite accurate estimate of the running couplings, without the
need to resort to `phenomenology' \cite{dv}.
For meson calculations with our choice of Hamiltonian (\ref{ham}),
additional phenomenological constraints must be used to obtain
unambiguous values for the coupling constants, although
symmetry requirements do strongly constrain them.

Of the other generators of the Poincar\'e algebra, $P^{\nu}, M^{\mu
\nu}$, the following can be derived canonically at $x^+ = 0$
\begin{eqnarray}
P^+ & = &  \int dx^- \sum_{{\bf x}, s, h} 2 \Tr  
                        \left\{ \partial_- M_s({\bf x})  
                \partial_- M_s({\bf x})^{\da} \right\}
+ {\rm i} u_{h}^{*} \partial_{-} u_{h}
 \ , \label{mom} \\
M^{-+} & = &   \int dx^- \sum_{{\bf x}, s, h}  x^- \left\{ 
             2 \Tr\left\{ \partial_{-} M_s({\bf x}) 
             \partial_{-} M_s({\bf x})^{\da} \right\} + {{\rm i}\over 2}  
u_{h}^{*} \stackrel{\leftrightarrow}{\partial}_{-} u_{h} \right\}  \ ,
   \\
M^{+r} & = & -  \int dx^- \sum_{{\bf x}, s, h} 
      2 \left( x^r + \frac{a}{2} \delta^{rs}\right) 
                      \Tr  \left\{ \partial_- M_s({\bf x})  
                \partial_- M_s({\bf x})^{\da} \right\} +   {\rm i} x^r 
u_{h}^{*} \partial_{-} u_{h} \ . \label{boost}
\end{eqnarray}
Note that these are all kinematic operators, quadratic in fields. 
$P^+$ and $M^{-+}$ generate translations and boosts respectively in
the $x^-$ direction and are unaffected by the transverse lattice
cut-off.
The cut-off effects on the boost-rotation operator $M^{+r}$ are
discussed further in the next section.

\subsection{Space of states}
\label{Fock}

For the construction of a Fock space of the dynamical fields $M_{r}$
and $u_{h}$, it is convenient to Fourier transform the fields in the 
$x^-$ co-ordinate only.
Thus, we introduce a  Fock space operator 
$a_{ r,ij}^{\dagger}(k^+,{\bf x})$ 
which creates a `link-parton' with light-cone momentum
$k^+$, carrying colour $i \in \{1, \cdots , N_c\}$ at site ${\bf x}$ to 
colour $j$ at site 
 ${\bf x} + a\widehat{\bf r}$; $a_{ -r,ij}^{\dagger}$ creates an
oppositely oriented link-parton. 
Likewise, $b_{h,i}^{*}(k^+,{\bf x})$ creates a `quark-parton' of helicity $h$, 
colour $i$,
momentum $k^+$ at site ${\bf x}$, while $d^*$ does the same for
anti-quarks. We have
\begin{eqnarray}
 \left[a_{\lambda,ij}(k^+,{\bf x}), 
        a_{\rho,kl}^{*}(\tilde{k}^+, {\bf y}\right] 
        & = & \delta_{ik}\, \delta_{jl}\, \delta_{\lambda \rho}\, 
        {\bf \delta_{x y}}\,\delta(k^+-\tilde{k}^+) \; , \\
   \left[a_{\lambda,ij}(k^+,{\bf x}),
        a_{\rho,kl}(\tilde{k}^+, {\bf y})\right] & = & 0 \; , \\
\left\{b_{h,i}(k^+,{\bf x}), 
        b_{h',j}^{*}(\tilde{k}^+, {\bf y})\right\} 
        & = & \delta_{ij}\,  \delta_{hh'}\, 
        {\bf \delta_{x y}}\,\delta(k^+-\tilde{k}^+) \;, \\
   \left\{b_{h,i}(k^+,{\bf x}),
        b_{h',j}(\tilde{k}^+, {\bf y})\right\} & = & 0 \; , 
\end{eqnarray}
where $\lambda$ and $\rho \in \{ \pm 1, \pm 2\}$ denote the
orientations of link variables in the $(x^1,x^2)$ plane, $a_{\lambda,ij}^{*} = 
a_{\lambda,ji}^{\dagger}$, and similar anti-commutators exist for $d$.
Fock space is already diagonal in the light-cone momentum
$P^+$ and serves as a basis
for finding the eigenfunctions $P^-$, the light-cone wavefunctions.
As usual in light-cone quantisation (without zero modes), 
the Fock vacuum state $|0\rangle$ is an exact eigenstate of $P^-$. 

Further cut-offs, apart from the transverse lattice, must be applied
to Fock space to make it finite-dimensional.
We will use  DLCQ \cite{dlcq,old} to
discretize light-cone momentum, which amounts to 
compactifying  $x^-$ on 
circle of circumference ${\cal L} = 2 \pi K/ P^+$, where $K$
is a positive  integer, with periodic (anti-periodic) boundary 
conditions for $M$ ($u$).  Eventually, we will
extrapolate observables to $K =\infty$. The use of anti-periodic boundary 
conditions is desirable because it tends to improve convergence as $K \to
\infty$. However, one cannot consistently have anti-periodic boundary 
conditions for both bosons and fermions in a theory with Yukawa-type
interactions.

To reduce the size of Fock space still further, it will be
convenient to impose a separate Tamm-Dancoff cut-off 
on the maximum number of partons in Fock space, studying 
the theory as this cutoff is raised. 
Since the large $N_c$ limit automatically restricts to a
quark-antiquark pair in the meson sector, this effectively means a
cut-off on the number of link-partons. 
A general meson state of  light-cone momentum $P^+$, which is
translationally invariant in the transverse direction,
takes the form
\begin{eqnarray}
&&|\psi(P^+)\rangle = \nonumber \\  
&& {2 a \sqrt{\pi} \over \sqrt{N_{c}} } \sum_{\bf x}  \sum_{h,h'}
\int_{0}^{P^+} dk^{+}_{1} dk^{+}_{2} \delta(P^+ -k^{+}_{1} -k^{+}_{2})
\left\{ \psi_{hh'}(x_1,x_2) 
\ b_{h}^{\dagger}(k^{+}_{1}, {\bf x})d_{h'}^\dagger(k^{+}_{2},
{\bf x}) |0\rangle \right\}\nonumber \\
&& + 
 {2 a \sqrt{\pi} \over N_{c} } \sum_{\bf x}  \sum_{h,h',r}
\int_{0}^{P^+} {dk_{1}^{+} dk_{2}^{+} dk_{3}^{+} \over P^+} 
\delta(P^+ -k^{+}_{1} -k^{+}_{2} -k_{3}^{+} ) \nonumber \\ 
&& \times \left\{ \psi_{h(r)h'}(x_1,x_2,x_3) 
 \  b_{h}^{\dagger}(k_{1}^{+}, {\bf x})a^{\dagger}_{r}(k_{2}^{+},{\bf x})
d_{h'}^*(k_{3}^{+}, {\bf x}+ a\hat{\bf r}) 
|0\rangle \right. \nonumber \\
&& \ \ \left. +  \ \ \psi_{h(-r)h'}(x_1,x_2,x_3)  
\  b_{h}^{\dagger}(k_{1}^{+}, {\bf x}+ a\hat{\bf r})
a^{\dagger}_{-r}(k_{2}^{+},{\bf x})
d_{h'}^*(k_{3}^{+}, {\bf x}) 
|0\rangle \right\} \nonumber \\
&& + 
 {2 a \sqrt{\pi} \over \sqrt{N_{c}^{3}} } \sum_{\bf x}  \sum_{h,h',r,s}
\int_{0}^{P^+} {dk_{1}^{+} dk_{2}^{+} dk_{3}^{+} dk_{4}^{+} \over (P^+)^2}  
 \delta(P^+ -k^{+}_{1} -k^{+}_{2} -k^{+}_{3} -k^{+}_{3})  \label{link}\\ 
&& \times \left\{ \psi_{h(rs)h'}(x_1,x_2,x_3,x_4) 
 b_{h}^{\dagger}(k_{1}^{+}, {\bf x})a^{\dagger}_{r}(k_{2}^{+},{\bf x})
a^{\dagger}_{s}(k_{3}^{+},{\bf x} +a\hat{\bf r})
d_{h'}^*(k_{4}^{+}, {\bf x}+ a\hat{\bf r}+ 
a\hat{\bf s}) 
|0\rangle \right. \nonumber \\
&& \left. +  \psi_{h(r-s)h'}(x_1,x_2,x_3,x_4) 
 \  b_{h}^{\dagger}(k_{1}^{+}, {\bf x})a^{\dagger}_{r}(k_{2}^{+},{\bf x})
a^{\dagger}_{-s}(k_{3}^{+},{\bf x} +a\hat{\bf r} -a\hat{\bf s})
d_{h'}^*(k_{4}^{+}, {\bf x}+ a\hat{\bf r}- 
a\hat{\bf s})
|0\rangle \right. \nonumber \\
&& \left. +  \psi_{h(-rs)h'}(x_1,x_2,x_3,x_4) 
 \  b_{h}^{\dagger}(k_{1}^{+}, {\bf x}+a\hat{\bf r})
a^{\dagger}_{-r}(k_{2}^{+},{\bf x})
a^{\dagger}_{s}(k_{3}^{+},{\bf x})
d_{h'}^*(k_{4}^{+}, {\bf x}+ a\hat{\bf s})
|0\rangle \right. \nonumber \\
&& \left. +  \psi_{h(-r-s)h'}(x_1,x_2,x_3,x_4) 
 \  b_{h}^{\dagger}(k_{1}^{+}, {\bf x}+a\hat{\bf r})
a^{\dagger}_{-r}(k_{2}^{+},{\bf x})
a^{\dagger}_{-s}(k_{3}^{+},{\bf x}-a\hat{\bf s})
d_{h'}^*(k_{4}^{+}, {\bf x}- a\hat{\bf s})
|0\rangle \right\}  + \cdots \nonumber \ ,
\end{eqnarray}
where states with up to two links have been explicitly displayed.
In (\ref{link}), $\dagger$ acts on gauge indices
and  $x_1 = k^{+}_{1}/P^+$ etc., are light-cone momentum fractions.
Only gauge singlet combinations under residual gauge transformations in
$A_{-}=0$ gauge can contribute to states of finite energy \cite{bard}. 
Because pair production of quarks and mixing with
glueballs is suppressed at 
large $N_c$, the states (\ref{link}) provide a description of the
valence quark content of flavour non-singlet mesons. Thus one should
implicitly  understand a distinct flavour label on the quark and
anti-quark, which is redundant. 
For states that are translationally invariant in the transverse direction, 
the transverse co-ordinate 
argument in wavefunctions may be
suppressed. The sequence of orientations $\lambda, \rho, \cdots$ of link
variables, together with the $P^+$ momentum fractions $x_1, x_2, \cdots$
and quark
helicities $h,h'$ are sufficient to encode the structure of Fock
states contributing to the boundstate. Thus, a general Fock state 
may be labelled
\be
| (x_1, h), (x_2, \lambda), \cdots, (x_{n-1},\rho), (x_n, h') \rangle \ .
\eq
The expansion (\ref{link}) may be  represented by a planar
(large-$N_c$) diagram
notation shown in Fig.~\ref{fig1}. 
This  will be helpful when enumerating the matrix
elements of the Hamiltonian.

The transverse momentum 
operator is not directly defined because of the lattice
regulator, but one may introduce transverse momentum ${\bf P}$ by  
application of the boost-rotation operator $M^{+r}$. Let 
$| (x_1, {\bf x_1}), \cdots, (x_{n},{\bf x}_{n}) \rangle$ 
denote an
$n$-parton Fock state.
${\bf x}_p$ is the transverse position and $x_p$ the $P^+$ momentum
fraction of the $p^{\rm th}$
parton (conventionally we take transverse position 
to be the midpoint of a link, for link
fields). Using ({\ref{boost}) we find
\be
 {\rm exp}\left[-{\rm i} M^{+r} P_{r}/P^+\right] | (x_1, {\bf x}_1), \cdots,
(x_{n},{\bf x}_{n}) \rangle
=   {\rm exp}\left[{\rm i} {\bf P}.\sum_{p=1}^{n} x_p {\bf x}_p\right]
| (x_1, {\bf x}_1), \cdots, (x_{n},{\bf x}_{n}) \rangle \ . 
\label{naive}
\eq
Therefore, the net effect is  to 
add phase factors  into matrix elements of $P^-$
between Fock states at ${\bf P}=0$.
In a Poincar\'e covariant or a free theory,  the transformation
(\ref{naive}) applied to eigenstates of $P^-$ (\ref{link}) 
would be sufficient to
generate eigenstates of $P^-$ at non-zero ${\bf P}$. However, the
lattice regulator spoils Poincar\'e covariance and in general one must 
rediagonalise $P^-$ after boosting Fock states by (\ref{naive}). Thus
the eigenfunctions $\psi$ in (\ref{link}) for $P^-$ 
will become functions of ${\bf P}$ also.

The state is  normalised covariantly 
\be
\langle \psi(P^{+}_{1},{\bf P}_{1})|\psi(P^{+}_{2},{\bf P}_2)\rangle =
2P^{+}_{1} 
(2\pi)^3 \delta(P^{+}_{1} - P^{+}_{2}) \delta({\bf P}_{1} - {\bf P}_{2})
\eq
if
\begin{eqnarray}
1  & = &  \int_{0}^{1} dx \sum_{h,h'} | \psi_{hh'}(x,1-x)|^2  
 + \int_{0}^{1} dx_1 dx_2 \sum_{h,\lambda,h'}  
|\psi_{h(\lambda)h'}(x_1,x_2,1-x_1-x_2)|^2 \nonumber \\
&& + \int_{0}^{1} dx_1 dx_2 dx_3\sum_{h,\lambda ,\rho,h'}  
|\psi_{h(\lambda \rho)h'}(x_1,x_2,x_3,1-x_1-x_2-x_3)|^2 + \cdots \ . 
\label{normed}
\end{eqnarray}
for any ${\bf P}_{1},{\bf P}_2$.
This also ensures that the light-cone momentum sum rule is
satisfied, even at finite DLCQ cutoff $K$, since translation invariance in the 
$x^-$ direction is preserved by DLCQ.  

Since there is $90^{\rm o}$ rotational symmetry about $x^3$ for a state
with ${\bf P}=0$, it is 
possible to distinguish the angular momentum projections
${\cal J}_3$ mod 4. There is also exact symmetry under 
${\cal G}$-parity, charge conjugation ${\cal C}$, and  transverse
reflections in the $x^1$ and $x^2$ direction, ${\cal P}_1$, ${\cal P}_{2}$. 
Although the parity ${\cal P} = {\cal P}_1 {\cal P}_{2} 
{\cal P}_{3}$ is dynamical and in general broken, one can associate a 
parity to boundstates from their behaviour under the free particle limit
of ${\cal P}_{3}$.   Indeed, 
there is a $Z_2$ kinematic symmetry 
\be
{\cal P}_f \psi_{hh'}(1-x,x) \to \psi_{hh'}(x,1-x)  \ ,  
\eq
which corresponds to the free ${\cal P}_{3}$ operation in the zero-link sector,
that is  exact  at any cut-off $K$. In this way, one has enough information
to identify the ${\cal J}^{{\cal P}{\cal C}}$ structure of light states
unambiguously.

In general, we will find that the $0^{-+}$ pion mass is 
split from the $1^{--}$ rho mass. 
%by the couplings $m_f k_a$ and $k_a k_s$ in table~\ref{table1}.
Due to violations of covariance, the ${\cal J}_3 = 0$ component of the rho
($\rho_0$) will also
split from its ${\cal J}_3 = \pm 1$ components ($\rho_{\pm}$) which are always
degenerate on the transverse lattice at ${\bf P}=0$. 
In view of the low-energy nature
of the truncation of the colour-dielectric expansion, we do not analyse
heavier mesons, although their eigenfunctions are obtained as a by product
of our calculations.

\subsection{Renormalisation}

We have constructed a gauge theory with transverse lattice and Tamm-Dancoff 
cut-offs that we do not intend to extrapolate and a DLCQ cut-off that
we do. The first step in the renormalisation process is to ensure
finiteness of physical observables in the limit $K \to \infty$. 
It turns out that divergences exist but they
require only infinite and finite self-energy  
counterterms that renormalise existing parton mass terms in the
light-cone Hamiltonian. The remaining cut-offs that are not
extrapolated obviously violate Lorentz covariance. This violation can 
however be minimized by appropriate finite renormalisation all of the
couplings appearing in $P^-$ (\ref{ham}). 
In this section we describe our procedure for performing these finite
and infinite renormalisations.

It is convenient to use
one of the parameters of the Hamiltonian to set the dimensionful scale
of the theory and define dimensionless versions of the
others. Conventionally we will use 
$\bar{G}$ to set the scale,  which has the dimensions of
mass. It will later be related to the QCD mass scale by calculation
of the heavy source potential \cite{mat1}.
The following dimensionless parameters are then introduced:
\begin{eqnarray}
&& {\mu_b \over \bar{G}} \to m_b \ \ ; \ \ {\mu_{f} \over \bar{G}} \to
m_f \ \ ;   {\kappa_a \sqrt{N_c} \over \bar{G}} \to k_a \ \ ; \ \
{\kappa_s \sqrt{N_c} \over \bar{G}} \to k_s   \ \ ;\nonumber \\ 
&& {\lambda_i \over  \bar{G}^2} \to l_i \ (i=1,2,4)
\ \ ; \ \
{\beta \over   \bar{G}^2} \to b \ . \label{coup} 
\end{eqnarray}

Since we will need to study the meson 
eigenfunctions of $P^-$ as a function of $P^+$ and
${\bf P}$, let us write, for these eigenfunctions,
\be
2P^+ P^- = {\cal M}^2 + R({\bf P}) \ .
\label{disperse}
\eq
such that $R(0) = 0$. ${\cal M}^2$ is the invariant mass (squared).
We begin with ${\bf P}=0$, in which case 
the non-zero Fock space matrix elements of the dimensionless
invariant mass operator
\be 
\langle (y_1, h_1), (y_2, \sigma), \cdots, (y_{n-1},\tau), (y_n, h_2) |
 2P^+ P^-/\bar{G}^2 | (x_1, h_1'), (x_2, \lambda), \cdots, 
(x_{n-1},\rho), (x_n, h_2') \rangle \label{nz}
\eq 
are enumerated in  figs.~\ref{fig2}(i)-(xiii) and table~\ref{table1}.
A number of comments are necessary to explain these amplitudes. 
We have defined
\be
{\rm Rot}[\lambda,\rho] \equiv  \epsilon_{|\lambda||\rho|} {\rm
Sgn}[\lambda] {\rm
Sgn}[\rho] \ . 
\eq
In the planar diagram vertices of Fig.~\ref{fig2}, light-cone momentum fraction
($x,y,z$), quark helicity ($h,h'$), and
link-field orientation ($\lambda,\rho,\sigma,\tau$) labels are given 
where necessary. Lines with a bar denote the $x^+$-instantaneous
propagators $\partial_{-}^{-1}$ and $\partial_{-}^{-2}$ for $v$ quarks and
$A_{+}$ gauge fields respectively.  `P' denotes
that a principal value prescription is used when integrating light-cone
continuum 
momentum fraction across singularities. For simplicity, we have not shown 
vertices involving only anti-quarks, which are similar to those involving
only quarks.
To these diagrams we add planar spectator lines which go to make up
the full gauge singlet Fock state. 

At finite transverse lattice spacing $a$, but before the light-cone 
DLCQ cutoff $K$
is imposed, the theory is behaving like
a continuum $1+1$-dimensional gauge theory coupled to a set of 
fundamental fermion
and adjoint scalar fields \cite{fran}. Although super-renormalisable in the 
$1+1$-dimensional sense, the light-cone quantisation in light-cone gauge
introduces
its own characterstic divergences due to the presence of non-local
instantaneous interations. 
Those originating from the instantaneous
gluon propagator $1/\partial_{-}^{2}$ are dealt with by the principal
value prescription in the manner established by 't Hooft \cite{hoof}. 
Those originating from the instantaneous
quark propagator $1/\partial_{-}$ have been studied by Burkardt \cite{mat4},
 whose
analysis we briefly recall. 

A basic one-loop logarithmic divergence
occurs in the quark self-energy as represented in the
Light-cone Perturbation Theory  
diagram of Fig.~\ref{fig3}(i) as the quark loop momentum vanishes.
The cubic vertices are of the same type, with
coupling either $m_f k_a$ or $m_f k_s$, once the orientations of the 
intermediate link fields have been summed over. The divergences 
are cancelled, in these diagrams and any others obtained by adding
spectators, by an infinite quark `kinetic' mass counterterm in the
Hamiltonian (Fig.~\ref{fig3}(ii))
\be
{(k_{a}^{2} +
k_{s}^{2}) \over \pi} \int_{0}^{x} {dy \over y} \ .
\eq
This is not sufficient for the divergences in the two-loop diagrams of 
Fig.~\ref{fig4}(i)(ii)(iii)
to cancel. One may add a finite kinetic mass counterterm $\delta m^2$, 
adjusted at order $(k_{a}^2,k_{s}^2)$, to produce finite results when 
Fig.~\ref{fig4}(iv) is included.  
Higher-loop generalisations of the same diagrams are also rendered finite
by adjusting $\delta m^2$ at higher orders in $k_a$ and $k_s$. Dressing loop
diagrams with instantaneous gluon lines (e.\ g.\ Fig.~\ref{fig5}) 
renders them individually 
finite.  As in Ref.~\cite{mat4}, our own checks of these statements for the
transverse lattice theory have been done only
in perturbation theory, but we will assume they are true to all orders.
It was also shown in Ref.~\cite{mat4}, by means of simple cases,
 that choosing the correct counterterm
$\delta m^2$ was equivalent to restoring parity invariance, which is not
manifest in light-cone co-ordinates.

By adjusting the finite counterterm $\delta m^2$, one ensures that the 
$K \to \infty$ limit can be taken when DLCQ is used. However, it was pointed
out by Burkardt that, while this ensures a finite answer for the 
$K \to \infty$ limit of the self-energy, the use of a momentum-independent
mass counterterm in DLCQ will not yield the 
same as the covariant answer for the same couplings. In effect DLCQ 
produces a finite violation of covariance. This is one of a number of sources
covariance violation in our calculation. Rather than analysing how one might
minimize the  individual violations --- it is not obvious which are the
most significant --- we will perform overall
covariance tests on the
boundstate wavefunctions that are the end-product.

A Tamm-Dancoff cut-off  on the maximum number of link fields in a 
state also violates covariance. 
In principle, this can be compensated by introducing spectator-dependent 
counterterms \cite{perry}. 
In practice, that will lead to too many couplings for
viable calculation at  
a physically reasonable choice of Tamm-Dancoff cut-off. However, it is 
necessary for finiteness of the quark self-energy to use spectator-dependent
finite counterterms $\delta m^2$. Therefore,  we must introduce separate
counterterms $\delta m^{2}_{p}$ for the Fock sector containing $p$ links
(note that the sector with $p$ maximum has no finite or infinite
quark self-energy counterterms). These are adjusted to produce
finite quark self energy
in addition to optimization of covariance of hadron wavefunctions. 
Since we work at the
level of hadrons, the quark self-energy is tested indirectly.
A tachyonic quark self-energy, whether divergent or not,
would be signalled by tachyonic behaviour in the lightest hadron
mass. 
Therefore, we test for  absence of such a divergence in the
pion mass as $K \to \infty$.
A positive divergent quark self-energy would artificially
suppress the lowest Fock sectors, that are subject to loop corrections
and counterterms, in the hadron wavefunction as $K \to \infty$
(the hadron mass may remain finite).
Therefore, we test for absence of this suppression, in particular by
fitting $f_\pi$ which is a measure of the $p=0$ Fock component.

In summary, taking the $K \to \infty$ limit with a Tamm-Dancoff
cut-off on link fields in place, one must introduce infinite and 
spectator-dependent finite self-energy counterterms. 
Even though the theory is now finite,
Poincare covariance is still violated
by the finite transverse lattice spacing 
$a$, the Tamm-Dancoff cutoff, and by the use of momentum-independent
finite-mass counterterms $\delta m_{p}^{2}$. 
We propose to minimize these violations by finitely
renormalising all the  couplings available in $P^-$.

For perfectly relativistic dispersion, 
$R({\bf P}) = |{\bf P}|^2$ for every eigenfunction in eq.~(\ref{disperse}); 
this will
receive corrections on the coarse transverse lattice.
To quantify the covariance violation we will expand
the dispersion relation for each boundstate 
\be
R({\bf P}) = c^2 |{\bf P}|^2 + O({\bf P}^4) \ .
\eq
The transverse speed of light $c$ will in general differ from one (the speed
in the $x^3$ direction). A simple criterion, which worked well
in previous studies, is to minimize this difference in the low
lying eigenstates of $P^-$, ignoring the anharmonic terms in $R$.

The same procedure may be carried out for glueball boundstates to
constrain the pure-glue interactions in the Hamiltonian, independently
of the meson sector (at large $N_c$). In addition,
the rotational invariance of the potential between heavy sources
may be optimized. These latter tests have been described in detail
in previous work \cite{dv}. We will use the string tension $\sqrt{\sigma}$
from the potential to set the QCD scale from experiment.

The transverse lattice lagrangian (\ref{ferlag}) contains terms that
also violate chiral symmetry explicitly, via the couplings $m_f$ and
$k_s$.
Since we work at the level of
hadrons, a measure of chiral current 
non-conservation is provided by the pion mass in a covariant stable theory,
as a result of the PCAC theorem.
This measure loses accuracy if the theory also has significant
explicit covariance violation, as in our case.
The explicit  violation of chiral symmetry 
could be minimized by tuning further chiral-symmetry violating counterterms,
which we discuss in the appendix.
However, these lie at higher order of the colour-dielectric expansion.
In the present calculation, we finitely 
renormalize the hamiltonian to fit the experimental pion mass, 
since this is naively 
a measure of chiral current conservation.
Since the explicit violation of chiral symmetry is actually larger than
that suggested by $\cal{M}_{\pi}$, 
we find that we must also fit the experimental rho mass
in order to maintain a realistic pi-rho splitting.

Thus, in addition to optimizing covariance via boundstate dispersion, we are
proposing to fit four experimental numbers 
$\cal{M}_{\pi}$, $\cal{M}_{\rho}$, $f_{\pi}$,
and $\sqrt{\sigma}$ in order to accurately determine the couplings in
our effective hamiltonian $P^-$. Since QCD with degenerate flavours
contains only two fundamental parameters, the transverse lattice
hamiltonian is not determined
from first principles. However, as described above, direct tests of parity 
and chiral symmetry might allow one to reduce the number of phenomenological
parameters further. We leave this for future work.

\section{Determination of Hamiltonian parameters}
\label{three}

In order to reduce the number of
coupling variables in the minimization process,
this is done in two stages. First, we examine 
glueball eigenfunctions of $P^{-}$, that contain only link-fields,
and the rotational invariance of the
groundstate potential between
two heavy sources of colour. Here we follow, with one exception, 
exactly the same procedure
used in refs.~\cite{dv} and so omit all details. 
The exception is that, instead of using
anti-periodic boundary conditions for link fields in $x^-$, we re-did
the calculations with periodic boundary conditions in order to be consistent
with the conditions used in the meson sector later. Note that these
`pure-glue' calculations extrapolate both $K$ and the Tamm-Dancoff cut-off,
constraining very precisely the couplings in $P^-$ relevant to that sector
($l_1,l_2,l_3,b,m_b$) when covariance is optimized. 
It is not necessary at this stage to use
any phenomenological input. 

We searched for a trajectory in
coupling space that optimized the Poincar\'e covariance of glueball
wavefunctions and the potential between heavy sources of colour.
A fairly well-defined one-parameter trajectory is picked out.
We chose the best point on that trajectory, which in effect fixes $a$, 
corresponding to a value of the link field mass $m_b =0.276$. 
At this point, we find $\bar{G} \approx 2.75 \sqrt{\sigma}$
and $a \bar{G} \approx 4$, where $\sigma$ is the string tension of the
asymptotically linear potential found between two heavy sources. 
If one takes $\sqrt{\sigma} = 440$ MeV, then 
$\bar{G} \approx 1200$ MeV and $a \approx  2/3$ fm.
The values of the other couplings determined by this point
are shown in table~\ref{table2}.

Having fixed a subset of the couplings, we fix the remaining ones sensitive
only to the meson sector. We investigated the Tamm Dancoff-cutoff up to four
links, but show results for a three-link cut-off, since a better sampling
of couplings is achievable in this case.
The transverse speed of light $c$ is optimized in the dispersion of the pion
and each component of the rho, together with the difference between the
calculated mass ${\cal M}_{\pi}$ and the physical value. As described
in the previous section,
we find we must include fits to the physical values of 
${\cal M}_{\rho}$ and $f_{\pi}$ in the 
optimization procedure in order to accurately
pin down the remaining
undetermined couplings of the Hamiltonian, which are shown in 
table~\ref{table2}. 

Table~\ref{table3}
 shows information on the pi and rho states at these couplings.
One notes that the spin $\pm 1$ projections of the rho still badly violate 
covariance, splitting the Lorentz multiplet and having asymmetrical
dispersion. 
Since the  rho is  
not yet behavingly covariantly overall, we do
not attempt a detailed phenomenological analysis of the resulting
wavefunctions.
On the other hand, we are able to achieve a relatively
symmetrical dispersion for the pion, with intercept ${\cal M}_{\pi}
=171$ MeV and 
decay constant $f_{\pi}=132$ MeV, close to the experimental values.
(The exact values would be  obtainable with a sufficiently fine sweep of 
couplings).

We checked that no Fock sectors are being artificially suppressed
and that the truncation to no more than three links is not causing
severe `finite-volume' effects. 
Table~\ref{table4}
 shows that
the peak in the transverse spatial distribution of the pion wavefunction is
well-accomodated by the three-link cut-off (results are similar for the
rho). However, the tail of the
wavefunction at four and higher links may contain a significant 
total probability,
which will affect observables sensitive to very small transverse momenta.
Therefore, in this paper we restrict our attention to observables 
integrated over all
available transverse momenta.
Indeed, when varying the Tamm-Dancoff cut-off above three links, we 
find very little change in the observables investigated
below.

\section{Pion Observables}
\label{four}

\subsection{Valence quark structure function}

The valence quark distribution function is defined as
\begin{eqnarray}
V(x) &  = &  \sum_{h, h'} |\psi_{hh'}(x,1-x)|^2 
 + \sum_{\lambda}\sum_{h,h'}
 \int_{0}^{1-x} dy  \  |\psi_{h(\lambda)h'}(x,y,1-x-y)|^2  \nonumber \\
&& + \sum_{\lambda , \rho}\sum_{h,h'}
 \int_{0}^{1-x} dy \int_{0}^{1-x-y} dz \  
|\psi_{h(\lambda \rho)h'}(x,y,z,1-x-y-z)|^2  + \cdots \ .
\end{eqnarray}
It is the probability for a quark to carry light-cone momentum fraction $x$.
The result we find for $V(x)$ in the pion 
on the transverse lattice in the three-link truncation
is shown in Fig.~\ref{fig6}. The raw (discrete) DLCQ data for $K=10,12,15,20$
are displayed together with an
extrapolation to $K \to \infty$. 
To produce this, at each K data is fit
to the form
\be
x V(x)=(1-x)^{\beta} x^{\alpha} (a +b\sqrt{x} + cx) \ . \label{mrs}
\eq
$xV$ is directly the momentum distribution and is easier to extrapolate
since it vanishes at $x=0$
We note that the simple form $x^{\alpha}(1-x)^{\beta}$,
used to parameterize early experimental data, is not sufficient to 
fit our result.
It is necessary to drop the $x=1/2K$ and $x=1-1/2K$ points from this analysis
since they do not join smoothly to the rest of the distribution.
This is because endpoint data
are prone to artifacts resulting from the vanishing of some of the
interactions in Table~\ref{table1}.
The smooth curves at each $K$ are then extrapolated pointwise, by a (good)
fit
to a quadratic in $1/K$, for a large set
of values in the interval  $0.1<x<0.9$. The grey region represents the
uncertainty from the extrapolation only.

The extrapolated data fits the form (\ref{mrs}) with $\beta=0.33(2)$,
$\alpha= 0.3(1)$, $a=0.33(3)$, $b=-1.1(2)$, $c=2.0(3)$. The errors are
from the extrapolation only. Bearing in mind that the extrapolation is based
on fits to data that do not cover the endpoint regions, the true errors on
$\alpha$ and $\beta$ are likely to be much larger.
From the first moment $\langle x V_{\pi} \rangle =
\int_{0}^{1} x V_{\pi}(x) dx$, we find that  $32\%$ of 
meson light-cone momentum is carried by
the quarks, with the same carried by the anti-quarks. 
In the range $0.1<x<0.9$, over which there is some measure of control, 
the result for $V_{\pi}$ is reminiscent of the constant $V_{\pi}=1$ 
distribution 
resulting from the chiral limit of chiral 
quark models (see Ref.~\cite{chiral} 
and references therein). However, because of the 
rapid rise at small $x$, which is expected on
general grounds from Regge-type behaviour \cite{ladder}, 
the flat part of the distribution
is at $V \sim 0.7$. Moreover, in the chiral quark models, all the light-cone
momentum is carried by quarks, while here $36\%$ is carried by the link
fields representing gluonic degrees of freedom.

A well-defined transverse scale is associated with $V(x)$ above, namely,
the transverse lattice spacing $a$. If we were to repeat the calculation
at a different $a$, one would expect to see an evolution of $V$ as a result
of the changing wavefunctions. In practice, the current
transverse lattice method
is only able to explore a small window in $a$ --- small enough to suppress
discretization errors but large enough for the use of massive disordered
link fields ---  which is too small
to reliably see evolution. Perturbative evolution equations typically use a 
different renormalisation scheme, so there is no simple 
matching between scales. 
This problem is common to most non-perturbative approaches
to QCD, except Euclidean lattice QCD where (moments of) $V$ can be
calculated in schemes matched to perturbation theory \cite{lattice}. 
In low-energy
effective theories for pion structure, 
such as QCD sum rules \cite{cz,nlcsr}, chiral quark, colour-dielectric
transverse lattice, truncated Dyson-Schwinger \cite{ds}, 
etc., one is reduced to
fixing the scale for input to perturbative evolution equations heuristically
by demanding, say, agreement of the first moment of $V$ with 
experiment at some scale.

If we demand that $\langle x V_{\pi} \rangle \approx 0.21$ at a scale
of 2 GeV, as suggested by the analysis of E615 and NA10
pion-nucleon Drell-Yan data
by Sutton et. al. \cite{sutt} for the
valence quark distribution, then 
the scale associated to our result, if it were used as input for 
leading order
non-singlet evolution, is $\mu \approx 500$MeV; this is 
reasonable given that $a^{-1} = 300$MeV.
In Fig.~\ref{fig7}
 we show our result for $x V_{\pi}(x)$ evolved to $6.6$GeV and 
compared with the raw data for the valence distribution deduced by
E615 \cite{e615} by combining data over scales $4-8.5$ GeV. For completeness,
 we
also show fits to $x V=x^{\alpha}(1-x)^{\beta}$  produced by earlier
experiments NA10 \cite{na10} and NA3 \cite{na3}. 
In the valence region $x > 0.5$, our
result agrees with the most recent experiment, which claims a more
accurate representation at large $x$. 
At smaller $x$ there is not much agreement, either between experiments
or with our result. This is hardly surprising given the sensitivity
of this region to assumptions about the sea quarks or their
measurement. In fact, our calculation
contains no sea quarks since it is at large $N_c$. The recently
discovered enhancement of initial-state interactions \cite{stan2}
is also expected to be most significant at small $x$, throwing
into doubt the simple connection between light-cone probabilities and
the Drell-Yan cross-section \cite{dy}. 
It is obviously desirable
to  have data on $V_{\pi}$ from sources other than the Drell-Yan process. 
This is also important from the theoretical perspective, given that
the current Dyson-Schwinger approach predicts a completely different shape
for the pion distribution function \cite{craig}.

\subsection{Distribution amplitude}
 
The distribution amplitude (in $A_{-} = 0$ gauge)
for the pion  is defined by
\begin{eqnarray}
\left. \left\langle 0\right| \,\overline{\Psi}(z)\, \gamma^{\mu} \gamma_5\, 
    \Psi(0)\,
   \left|\psi_{\pi}(P^{\mu})\right\rangle \right|_{z^2 = 0} & = & 
   f_{\pi} P^{\mu} \int_{0}^{1}
{\rm e}^{{\rm i} x (z.P)} \phi_{\pi}(x) dx \label{pidist} 
\end{eqnarray}
with the normalisation condition 
\be 
\int_{0}^{1} \phi_{\pi}(x) = 1 \ .
\eq
If the quark field correlator is to be evaluated at equal light-cone
time, $z^+ = 0$, then ${\bf z} = 0$ and $z^-$ is arbitrary. This then
measures the amplitude for zero transverse separation of quarks in the
meson light-cone wavefunction. 
For the transverse lattice one finds
\be
\psi_{+-}(x,1-x) =   {f_{\pi} \over 2} \sqrt{\pi \over N_c}  \phi_{\pi}(x)  
\eq
from the $\gamma^+$ component of (\ref{pidist}). 
Fig.~\ref{fig8} shows our results for the distribution amplitude  at various
 $K$ and extrapolated to $K = \infty$ in a three-link truncation.  
The raw (discrete) DLCQ data has been  fit at each $K$ to the first few terms
of the conformal
expansion \cite{conf1,conf2} 
\be
\phi_{\pi}(x)  =  
6 x (1-x) \left\{ 1 + a_2 C_{2}^{3/2}(1-2x) + a_4 C_{4}^{3/2}(1-2x)
\right\} 
 \ . \label{exppi}
\eq
An extrapolation of the coefficients with $A + B/K + C/K^2$ 
yields $a_2 = 0.15(2)$, $a_4=0.04(1)$, confirming that truncation of the
conformal expansion is justified.
The same values are obtained if the fit curves at each $K$ 
are pointwise extrapolated
and then refit to (\ref{exppi}). Finally, the result is insensitive to whether
the $x=1/2K$ and $x=1-1/2K$ endpoints are included in the fit or not, so we
have shown them in Fig.~\ref{fig8} also.

The distribution amplitude is  indirectly
accessible through the pion transition form factor 
$F_{\pi \gamma^* \gamma}(Q^2)$ 
measured at CLEO \cite{cleo}.
A perturbative QCD analysis relates this to the inverse moment, up to
radiative corrections $\Delta$,
\be
{3Q^2 \over 4\pi} F_{\gamma^* \gamma \pi} = 
\int_{0}^{1} {\phi_{\pi}(x) \over x} + \Delta  = 3(1 + a_2 + a_4) + \Delta  
\eq
An analysis of the data in Ref.~\cite{sk}
extracted $a_2 + a_4 = 0.05\pm 0.07$
at scale 2.4 GeV, taking into account next-to-leading order 
corrections $O(\alpha_s)$ in $\Delta$. If we assume, following the structure
function analysis,  a scale $0.5$GeV for the transverse lattice result,
when evolved to 2.4 GeV by the 1-loop evolution equations we find
$a_2 = 0.07(1)$, $a_4 = 0.01(1)$ including only DLCQ errors. 
Although our result seems consistent with experiment, a  couple of 
comments are necessary. The inverse moment
is highly sensitive to the endpoint regions of $\phi_{\pi}$, which are
not well covered by the extrapolation of the DLCQ transverse lattice result.
Also, the leading radiative corrections  
in $\Delta$ are large $\sim 20 \%$, so one might ask about  higher order 
corrections. The reader is referred to refs.\cite{chiral,bakulev}
 for a more detailed review
of the various theoretical and experimental results relating to 
$\phi_{\pi}$. 

Diffractive dissociation on a nucleus $\pi + A \to A + {\rm
jets}$ \cite{ashery} has been used to measure a cross-section
related to  $\phi_{\pi}$. A number a theoretical analyses of that relation
have recently been performed \cite{diss}, which 
differ in their conclusions about the precise relationship. 
Our result,
when evolved to the higher transverse momentum scale of the experiments,
is consistent with any one of the analyses, being close to the asymptotic
form $6 x (1-x)$. 
We mention that our DLCQ transverse lattice 
result for $\phi_{\pi}$ is close  to one
previously obtained in a one-link truncation using very similar methods
\cite{sd}, 
although $a_4$ was not fit
and the normalisation $f_{\pi}$ was completely wrong in that case. The
same 1-link truncation was investigated in Ref.~\cite{mat3} by using
basis functions instead of DLCQ and a similar (but not identical)
criteria for fixing the
Hamiltonian couplings. That gave a distribution amplitude  a little closer
to the asymptotic form, although a value for $a_2$ was not extracted and
no error estimate was given.
Thus, we can say
with some confidence that our result is neither the `double-hump' first
found by Chernyak and Zhitnitsky \cite{cz} using (local) sum rules
  nor the `narrow-hump' one would deduce
from the latest Euclidean lattice measurements of the lowest moment of
$\phi_{\pi}$ \cite{lat}. 

\subsection{Quark helicity correlation}

Although the pion is spin 0, it nevertheless contains a complicated
spin structure. One measure of this is the quark helicity correlations
\begin{eqnarray}
C^{\rm para}(x) &  = &  \sum_{h} |\psi_{hh}(x,1-x)|^2 
 + \sum_{\lambda}\sum_{h}
 \int_{0}^{1-x} dy  \  |\psi_{h(\lambda)h}(x,y,1-x-y)|^2  \nonumber \\
&& + \sum_{\lambda , \rho}\sum_{h}
 \int_{0}^{1-x} dy \int_{0}^{1-x-y} dz \  
|\psi_{h(\lambda \rho)h}(x,y,z,1-x-y-z)|^2  + \cdots \ , \\
C^{\rm anti}(x) &  = &  \sum_{h} |\psi_{-hh}(x,1-x)|^2 
 + \sum_{\lambda}\sum_{h}
 \int_{0}^{1-x} dy  \  |\psi_{-h(\lambda)h}(x,y,1-x-y)|^2  \nonumber \\
&& + \sum_{\lambda , \rho}\sum_{h}
 \int_{0}^{1-x} dy \int_{0}^{1-x-y} dz \  
|\psi_{-h(\lambda \rho)h}(x,y,z,1-x-y-z)|^2  + \cdots \ . 
\end{eqnarray}
They measure the probability for a quark to have light-cone momentum
fraction $x$ and helicity either parallel or anti-parallel to that of the 
anti-quark. Therefore their sum is normalised to one (when integrated over 
$x$).
These functions are plotted in Fig.~\ref{fig9} for the
pion. Although 
different from one another in the bulk of $x$, both $C^{\rm para}_{\pi}(x)$ 
and $C^{\rm anti}_{\pi}(x)$ 
have exponents $\alpha$ and $\beta$ consistent within
errors with those of $V_{\pi}(x)$ (see Fig.~\ref{fig6}). We estimate
that 
\be
\int_{0}^{1} C^{\rm para}_{\pi} dx \sim 0.45 \ .
\eq
Therefore, one is almost
equally likely to find quark helicities aligned as anti-aligned!

\section{Conclusions}

We have extended coarse transverse lattice calculations to physically
realistic cut-offs on the anti-quark---quark separation. A general 
light-cone
hamiltonian in the large $N_c$ limit 
was expanded in powers of dynamical fields and we studied
a truncation of that colour-dielectric expansion. This included all 
possible cubic terms
and most of the quartic terms. By optimising
Lorentz covariance of glueball, heavy-source and 
meson boundstates, the remaining
freedom in the couplings in the Hamiltonian was reduced. By studying
other symmetries, such as parity and chirality, it may be possible to
constrain them further.  In this paper, 
we performed a phenomenological calculation
by fixing the remaining freedom in the couplings to best fit
$\sqrt{\sigma}$, ${\cal M}_{\pi}$, ${\cal M}_{\rho}$, and $f_{\pi}$
(two of these are parameters of first-principles QCD). 

The lightest meson bound state has the quantum numbers of the pion
and exhibits a reasonably covariant lightcone wavefunction. Comparing
the predictions of this  wavefunction
with various experimentally measured observables for the pion, we find
consistency in the regions insensitive to sea quarks. New observables,
which in principle can be extracted from a higher twist analysis
of experiments, follow from the multi-parton correlations in the
light-cone wavefunction. As an example, we computed the anti-quark---quark
helicity correlation, with somewhat surprising results. Because
the tail of the pion wavefunction is still truncated in our calculation,
we did not compute observables sensitive to small transverse momentum.
Nevertheless, it would be interesting to look at the  general features
of the skewed parton distributions for intermediate momentum 
transfers, since little hard information is available for these
important observables.
It should be straightforward to extend the calculations to
strange mesons and heavy-light mesons.

There are still some shortcomings in the calculation.
The rho boundstate is not yet behaving covariantly. Our 
optimization of chiral symmetry
could be considerably improved. 
Given the close connection of Lorentz and chiral
symmetry on the lattice, we believe that these problems are related. 
In particular, higher-order terms in the
colour-dielectric expansion can fulfill a dual role to improve
both these symmetries.

\vspace{10mm}
\noindent {Acknowledgements}:
The work of SD was supported in part by PPARC grant GR/LO3965. 
BvdS was supported by an award from Research Corporation.
We would like to thank Geneva College undergraduates E. M. Watson
and J. Bratt for help with developing the numerical code.

\vspace{10mm}
\noindent {\bf APPENDIX: Chiral symmetry}

\vspace{10mm}
The lattice Lagrangian (\ref{ferlag}) explicitly  breaks
chiral symmetry 
\be
\Psi \to e^{-{\rm i}  \theta \gamma_5}\Psi \ , \label{trans}
\eq
through the bare mass-term  $\mu_f$ and Wilson term $\kappa_s$.
The standard test for this at the hadron level is PCAC
\be
\langle 0| \partial_{\mu} A^{\mu} |\psi_{\pi}\rangle
= f_{\pi}
{\cal M}_{\pi}^{2} \ ,
\label{pcac}
\eq
where $A_{\mu}$ is the axial current. Without knowing the 
precise form of $A_{\mu}$, one can use ${\cal M}_{\pi}$ to quantify
the amount of explicit chiral symmetry breaking relative to other
scales, such as the pure-QCD mass gap or the spontaneous chiral symmetry
breaking scale given by the difference between ${\cal M}_{\pi}$ and
masses of other light mesons. 
The result (\ref{pcac}) relies on exact Lorentz covariance, which is not
present on the transverse lattice. In fact, in the calculation performed
in this paper, the splitting of the $\rho$ Lorentz multiplet is of comparable
strength to the $\pi$-$\rho$ splitting. This suggests that explicit
chiral symmetry breaking effects are larger than ${\cal M}_{\pi}$
would suggest, perhaps of the same order as spontaneous chiral symmetry
breaking effects.

Explicit chiral symmetry breaking could in principle be tested more
directly.
There is a $1+1$-dimensional Noether `vector' current 
\be
j^{\alpha}= \sum_{\bf x}
\overline{\Psi}({\bf x}) \gamma^{\alpha} \Psi({\bf x})
\eq
that is conserved under the equations of motion, 
i.e. $\partial_{\alpha} j^{\alpha} = 0$, and a corresponding chiral
current
\be
j_{5}^{\alpha}= \sum_{\bf x}
\overline{\Psi}({\bf x}) \gamma^{\alpha} \gamma_5 \Psi({\bf x})
\eq
for which we find
\be
\partial_{\alpha} j_{5}^{\alpha} =  2 \mu_{f}^{2} \sum _{{\bf x}, h}
h F_{h}^{\dagger}({\bf x})  {1 \over \partial_{-}} F_{h} ({\bf x})  \ .
\label{chiral}
\eq
where $F_{h}$ is defined in eq.~(\ref{eff}).
One might then use a  matrix element such as
\be
\langle 0| \partial_{\alpha} j_{5}^{\alpha} |\psi_{\pi}\rangle
\label{div}
\eq
to quantify explicit chiral symmetry violation, minimizing
it by finite renormalisations of couplings, since the vanishing of 
(\ref{div}) is a necessary 
condition for conservation of the four-dimensional axial
current. There  are a few difficulties
that must be overcome before this would be practical however. The expression
(\ref{chiral}) has a normal-ordering ambiguity similar to the hamiltonian.
Moreover, it is much more computationally expensive to perform symmetry
tests with eigenfunctions rather than eigenvalues. On a coarse lattice,
the chiral 
symmetry breaking couplings are also strongly constrained away from zero
by
Lorentz covariance; for example, $\kappa_s$ is needed to avoid
fermion doubling \cite{mat2}. It would therefore be desirable to have
further independent chiral symmetry breaking couplings in the theory to
tune.

Natural candidates are the transverse lattice versions of the
Sheikholeslami-Wohlert terms \cite{csw}, $\overline{\Psi} \sigma^{\mu \nu}
F_{\mu \nu} \Psi$,
that can be used in Euclidean
lattice gauge theory to remove $O(a)$ contributions to chiral 
current non-conservation \cite{lusher}. They become
\be
\overline{\Psi}({\bf x}) \sigma^{rs}\left( M_{r}({\bf x})M_{s}({\bf x}+a 
\hat{\bf r})  - M_{s}({\bf x})M_{r}({\bf x}+a \hat{\bf s}) \right)
\Psi({\bf x}+ a \hat{\bf r}+a \hat{\bf s} ) \ , \label{one}
\eq
\be
\overline{\Psi}({\bf x}) \sigma^{+-} F_{+-} \Psi({\bf x}) \label{twice}
\eq
While in the dimensional counting classification of 
Euclidean lattice operators
they occur along with Wilson terms at dimension five, 
on the coarse transverse lattice their significance is not
so obvious since they contribute to higher orders of the
colour-dielectric expansion. 
Terms of the form (\ref{one})(\ref{twice}) in the
Lagrangian give rise to coupled constraint equations of motion for
non-dynamical fields. If solved order by order in dynamical fields, they
give rise to new interactions in the gauge-fixed light-cone hamiltonian
starting at order $u^4$, $u^2 M^2$. Of particular interest are
interactions generated at order $u^2 M^3$ and $u^4 M$ that flip the
helicity $h$ of quarks. Such interactions carry the
spontaneous chiral symmetry breaking effects in effective light-cone
hamiltonians \cite{wilson1}; 
the $m_f k_a$ and $k_s k_a$ terms performed this
function in (\ref{nz}).

\newpage

\begin{table}
\centering$\displaystyle
\renewcommand{\arraystretch}{1.25}
\begin{array}{|c|c|}
\hline
{1 \over x}\left(m_{f}^{2} + \delta m_{p}^{2} + {(k_{a}^{2} +
k_{s}^{2}) \over \pi} \int_{0}^{x} {dy \over y} \right)& (i)  \\
\hline
{1\over 2\sqrt{\pi y}} \left({1\over x+y} -
{1\over x}\right)\left\{m_f k_s
+ m_f k_a {\rm Sgn}[\lambda] (\delta_{|\lambda|2} - 
{\rm i}h\delta_{|\lambda|1})\right\} & (ii) \\
\hline
{-1 \over 2 \pi} \ {\rm P}\left( {1 \over (x-y)^2} \right) & (iii) \\
\hline
{-(y-z)  \over 4 \pi (y+z)^2 \sqrt{y z}} \delta_{-\lambda \rho} & (iv) \\
\hline
{1 \over 4 \pi (x+y)\sqrt{yz}}\left\{ k_{s}^{2} \delta_{hh'} +
k_{a}^{2} \delta_{hh'}\left(\delta_{\lambda \rho} - \delta_{-\lambda \rho}
-{\rm i}h{\rm Rot}[\lambda, \rho]\right) \right. & (v)  \\
\left. + k_a k_s \delta_{-hh'} 
\left({\rm Sgn}[\lambda]({\rm i}h\delta_{|\lambda|1}
-\delta_{|\lambda|2}) + {\rm Sgn}[\rho]({\rm i}h\delta_{|\rho|1}
-\delta_{|\rho|2})\right)\right\} & \\
\hline
{1 \over 4 \pi (x+y)\sqrt{yz}}\left\{ k_{s}^{2} \delta_{hh'} +
k_{a}^{2} \delta_{hh'}\left(\delta_{-\lambda \rho} - \delta_{\lambda \rho}
+{\rm i}h{\rm Rot}[\lambda, \rho]\right) \right. & (vi)  \\
\left. 
+ k_a k_s \delta_{-hh'} \left({\rm Sgn}[\lambda]({\rm i}h\delta_{|\lambda|1}
-\delta_{|\lambda|2}) - {\rm Sgn}[\rho]({\rm i}h\delta_{|\rho|1}
-\delta_{|\rho|2})\right)\right\}\nonumber &  \\ 
\hline
{m_{b}^{2} \over x} & (vii) \\
\hline
{-1 \over 4 \pi} {\rm P}\left( {y+z \over \sqrt{z y} (z-y)^2}
\right) & (viii) \\
\hline
{-1 \over 8 \pi} {\rm P}\left( {(y+z)(2x+y-z) \over \sqrt{xzy(x+y-z)} 
(z-y)^2} \right) & (ix) \\
\hline
{1 \over 4 \pi \sqrt{xyz(x+z-y)}} \left\{ 2l_1 \delta_{\sigma \lambda}
\delta_{\rho \tau} \delta_{-\rho \sigma} +  l_2 ( \delta_{\sigma \lambda}
\delta_{\rho \tau} \delta_{\rho \sigma} + \delta_{-\sigma \rho}
\delta_{-\lambda \tau} \delta_{-\sigma \lambda}) \right. & (x) \\ 
 \left. +l_4 (\delta_{\sigma \lambda}
\delta_{\rho \tau}|{\rm Rot}[\sigma,\rho]| +\delta_{-\sigma \rho}
\delta_{-\lambda \tau}|{\rm Rot}[\sigma,\lambda]|)
-b\delta_{\lambda \rho} \delta_{\sigma \tau}|{\rm
Rot}[\sigma,\rho]|\right\} & \\
\hline
{1 \over 4 \pi \sqrt{xyz(x+z+y)}}
 \left\{ 2l_1 \delta_{-\sigma \lambda}
\delta_{\lambda \tau} \delta_{\lambda \rho } +  l_2  \delta_{\sigma \tau}
\delta_{-\lambda \rho} \delta_{|\lambda| |\sigma|} \right. & (xi) \\
 \left. + l_4 
(\delta_{-\lambda \sigma} \delta_{\rho \tau}|{\rm Rot}[\lambda,\rho]|
+ \delta_{-\sigma \rho}
\delta_{\lambda \tau}|{\rm Rot}[\sigma,\lambda]|)
-b\delta_{-\lambda \rho} \delta_{\sigma \tau}|{\rm
Rot}[\sigma,\lambda]|
\right\} & \\
\hline
{-1 \over 8 \pi} {(y-z)(2x+y+z) \over \sqrt{xzy(x+y+z)} (z+y)^2}
\delta_{-\lambda \rho} & (xii) \\
\hline
{-1 \over 8 \pi} {(x-z)(2y-x-z) \over \sqrt{xzy(x-y+z)} (z+x)^2}
\delta_{-\sigma \rho}\delta_{-\lambda \tau} & (xiii) \\
\hline
\end{array}$
\caption{Matrix elements of the dimensionless invariant mass operator
$2P^+P^-/\bar{G}^2$ in Fock space. Momentum conserving delta functions
are omitted for clarity.
\label{table1}}
\end{table}

\begin{table}
\centering$\displaystyle
\renewcommand{\arraystretch}{1.25}
\begin{array}{|c|c|c|c|c|c|c|c|c|c|c|}
\hline
m_b & b & l_1 & l_2 & l_4  & k_s & k_a & m_{f} & \delta m_{0}^{2} & 
\delta m_{1}^{2} & \delta m_{2}^{2} \\ 
\hline 0.276 & 0.768 &  -0.169 & -0.186 &  0.024 &
0.420 & 0.652 & 0.236 & 1 & -0.127 & -0.035 \\
\hline
\end{array}$
\caption{Optimum coupling constants at $a= 2/3$ fm for a three-link 
truncation. Note that 
$\delta m_{0}^{2}$ was swept more coarsely than the other couplings.
\label{table2}}
\end{table}

\begin{table}
\centering$\displaystyle
\renewcommand{\arraystretch}{1.25}
\begin{array}{|c|c|c|c|c|c|c|c|c|c|c|}
\hline 
{\rm State} & {\rm Mass (MeV)} & c \\
\hline 
\pi & 171 & 1.02 \\
\hline
\rho_0 & 828  & 0.99 \\
\hline
\rho_+ & 457 & 1.04 \\
\hline
\rho_- & 457 & 0.76 \\
\hline
\end{array}$
\caption{Meson dispersion at the optimum couplings.
\label{table3}}
\end{table}

\begin{table}
\centering$\displaystyle
\renewcommand{\arraystretch}{1.25}
\begin{array}{|c|c|c|c|c|c|c|c|c|c|c|}
\hline
\# {\rm Links} &  0 &  1 & 2 & 3   \\ 
\hline {\rm Probability} & 0.097(14) & 0.661(10) & 0.150(8) & 0.087(2)  \\
\hline
\end{array}$
\caption{Probability for finding a certain number of links in the pion.
The extrapolation errors in brackets are from a $1/K$ extrapolation. 
\label{table4}}
\end{table}

\newpage

\begin{figure}
\centering
\BoxedEPSF{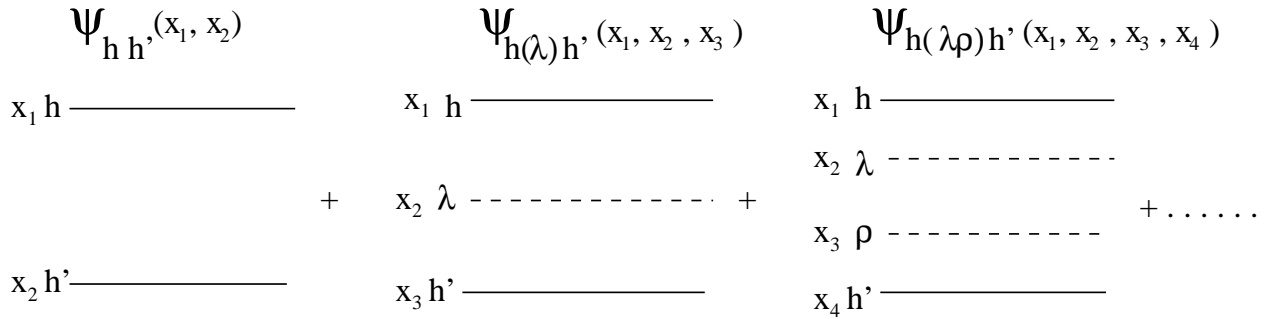 scaled 700}
\caption{ Planar diagram representation of the Fock space structure
of a meson boundstate. Solid lines represent quarks/anti-quarks, chain
lines link fields. \label{fig1}}
\end{figure}

\newpage

\begin{figure}
\centering
\BoxedEPSF{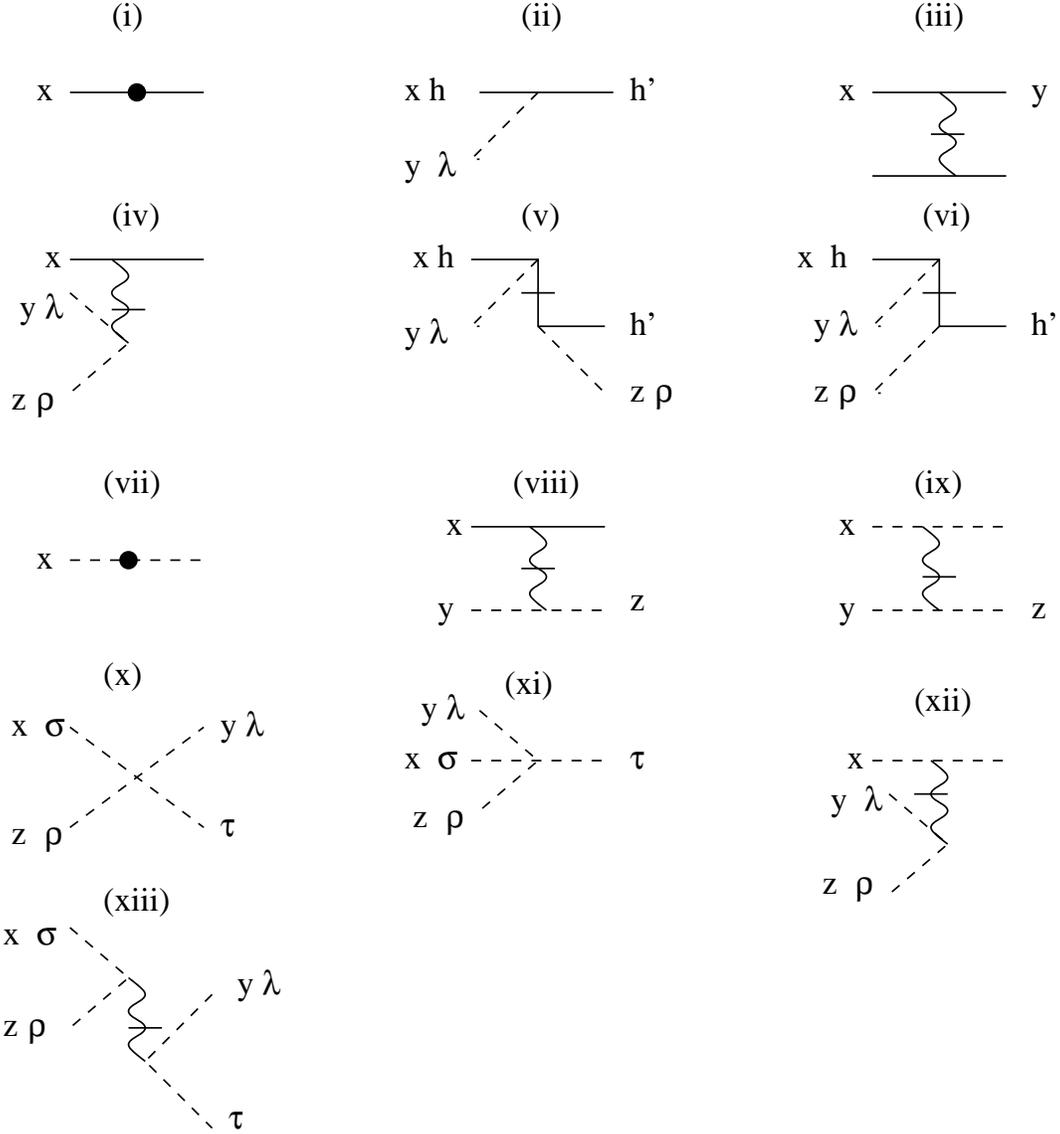 scaled 700}
\caption{ Planar diagram representation of the elementary amplitudes
contributing to (\ref{nz}). Vertical barred lines are $x^+$-instantaneous
interactions. \label{fig2}}
\end{figure}

\begin{figure}
\centering
\BoxedEPSF{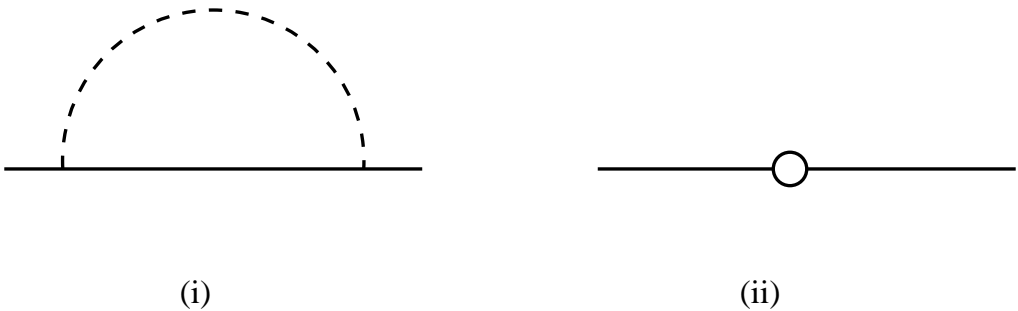 scaled 700}
\caption{ (i) one-loop logarithmically divergent quark self-energy
(ii) logarithmically divergent mass insertion counterterm represented by 
open circle .\label{fig3}}
\end{figure}

\begin{figure}
\centering
\BoxedEPSF{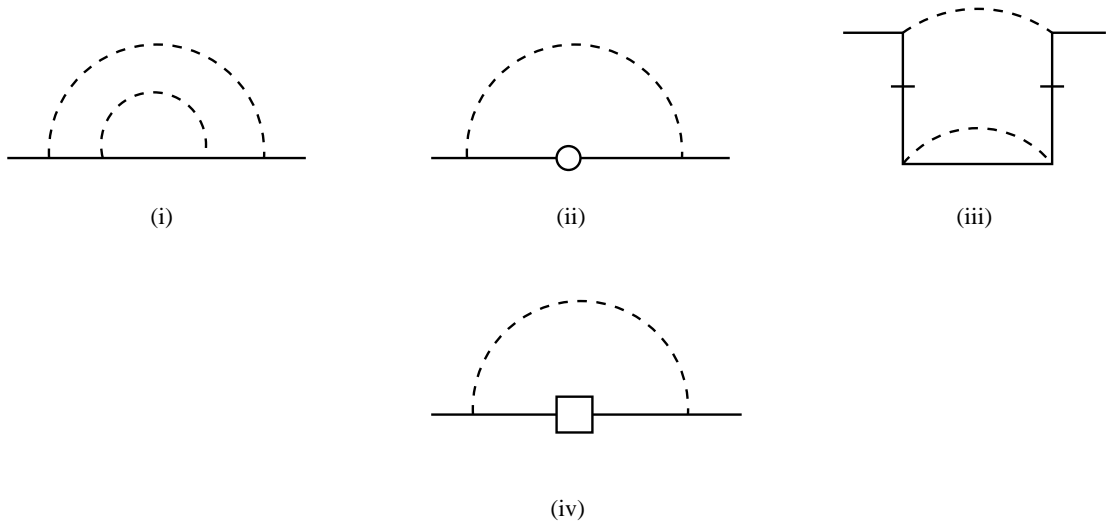 scaled 500}
\caption{ (i)(iii) two-loop logarithmically divergent quark
self-energies (ii) one-loop diagram with infinite mass insertion (iv)
one-loop diagram with finite mass insertion $\delta m^2$ represented
by open box\label{fig4}}
\end{figure}

\begin{figure}
\centering
\BoxedEPSF{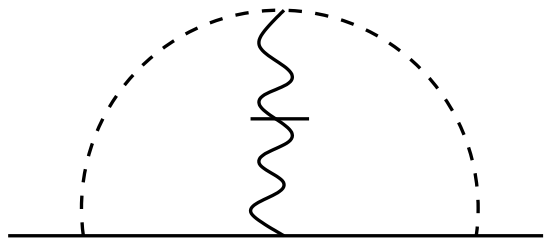 scaled 700}
\caption{ Finite diagram with instantaneous interaction dressing \label{fig5}}
\end{figure}

\begin{figure}
\centering
$\displaystyle V_{\pi}$ \hspace{5pt}  \BoxedEPSF{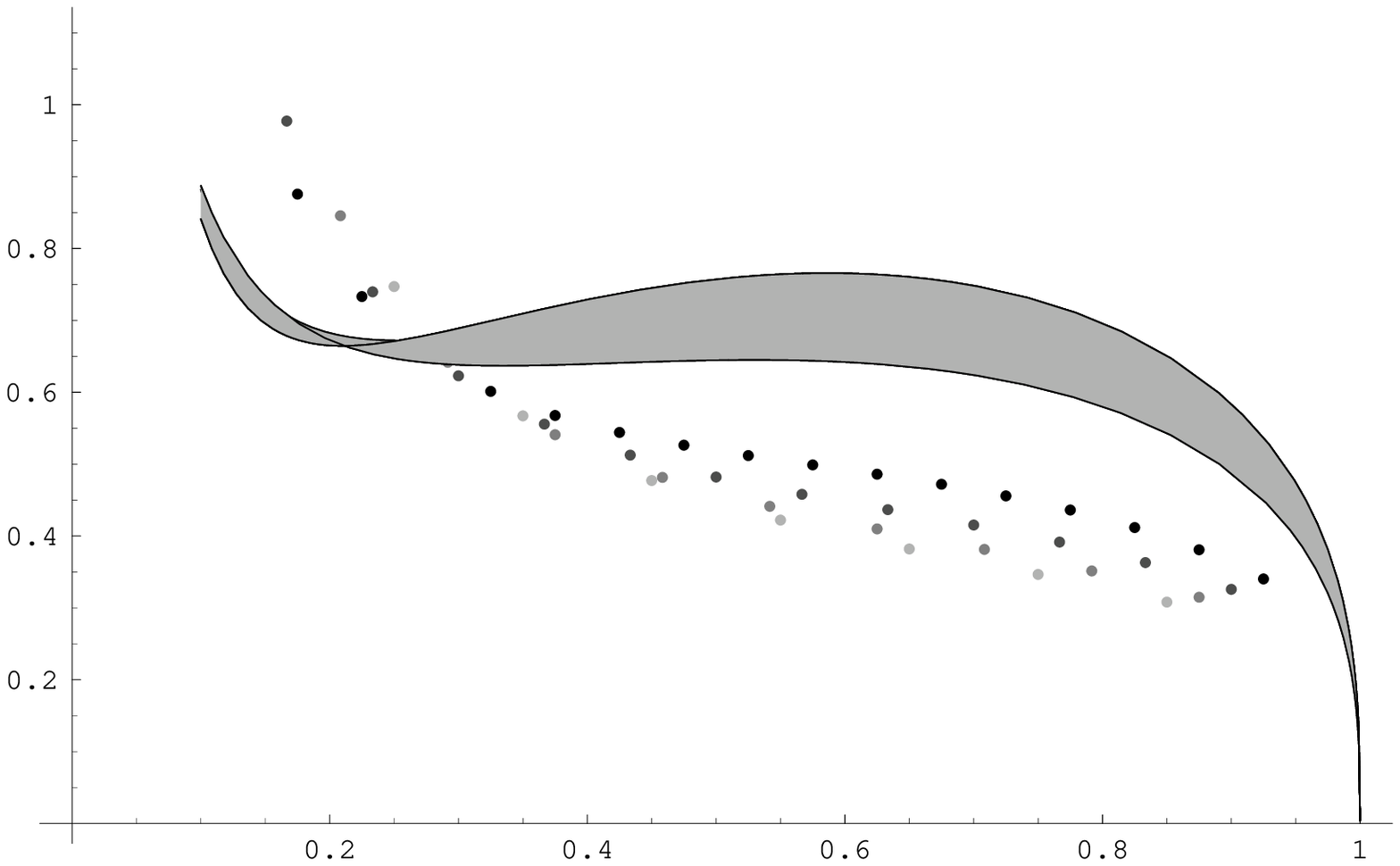 scaled 800}\\
\vspace{-1.7in} \hspace{0.5in} $\displaystyle x$
\vspace{0.5in}
\caption{ Pion distribution function for DLCQ cutoffs $K=10,12,15,20$, 
darker data points meaning larger $K$. $K \to \infty$ extrapolated curve
lies in the shaded region.  
\label{fig6}}
\end{figure}

\begin{figure}
\centering
\BoxedEPSF{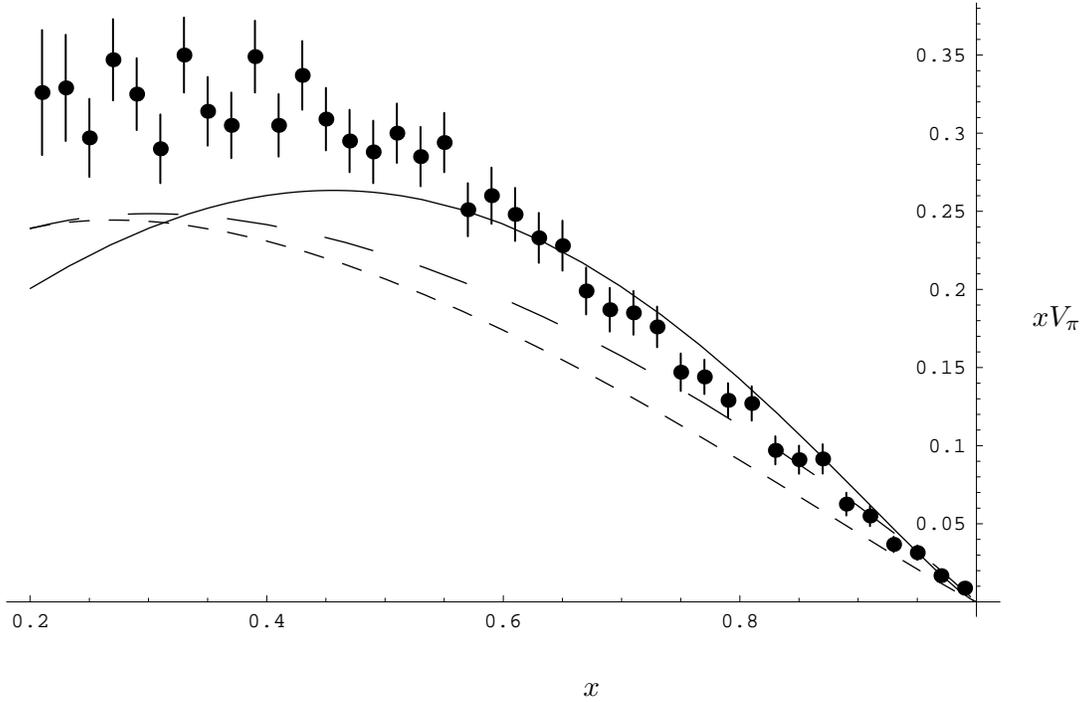 scaled 800} \hspace{5pt} $\displaystyle x V_{\pi}$ \\
\vspace{-1.7in} \hspace{0.5in} $\displaystyle x$
\vspace{0.5in}
\caption{ Pion (valence) distribution function (times $x$) compared
to pion-nucleon Drell-Yan data. 
Solid line: transverse
lattice result evolved to $6.6$ GeV. Data points: E615 experiment.
Short-dashed line: NA10 experiment fit to $x^{\alpha} (1-x)^{\beta}$ form.
Long-dashed line: NA3 experiment  fit to $x^{\alpha} (1-x)^{\beta}$ form.
\label{fig7}}
\end{figure}

\begin{figure}
\centering
$\phi_{\pi}$  \hspace{5pt} \BoxedEPSF{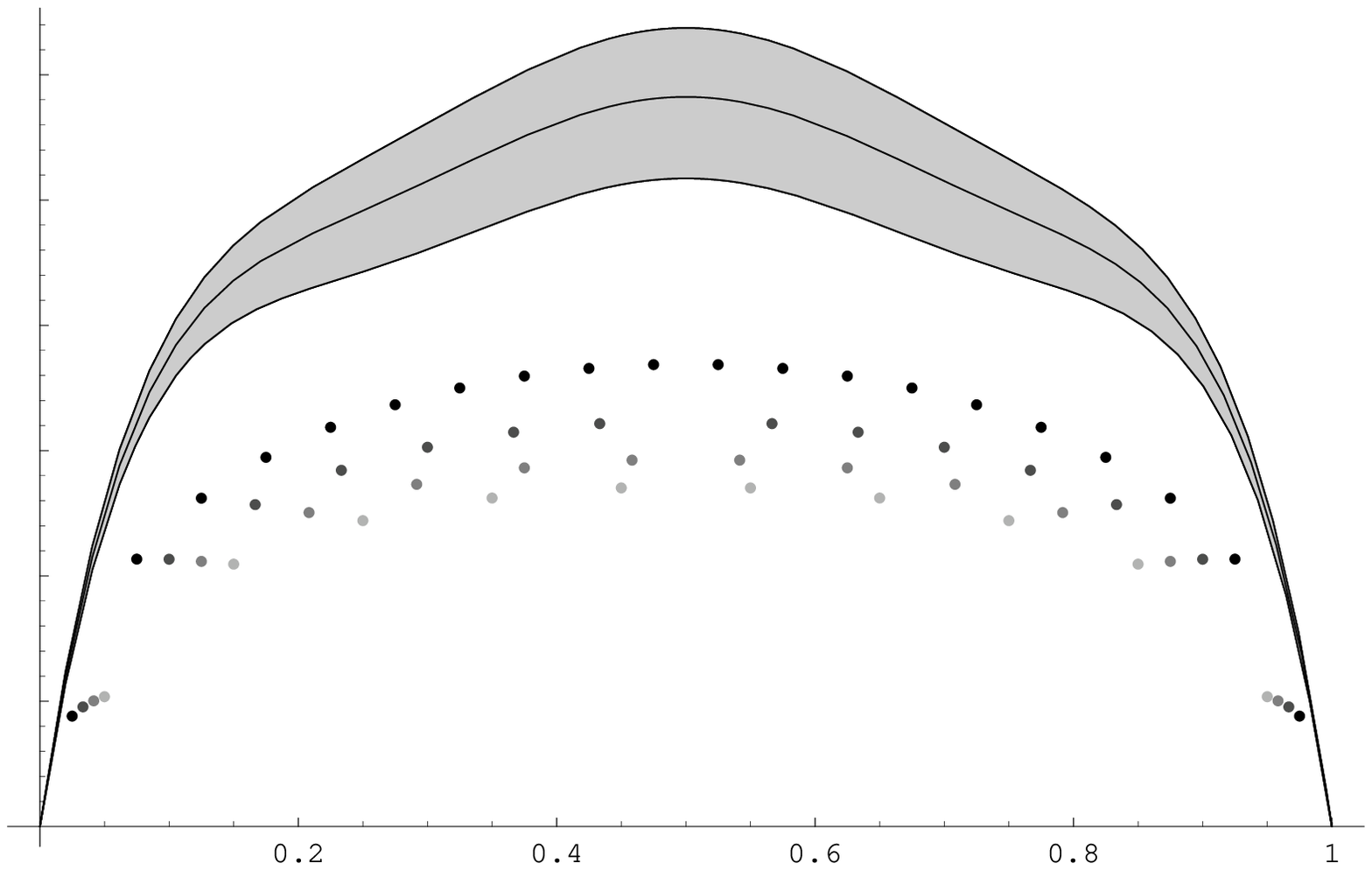 scaled 800}   \\
\vspace{-1.7in} \hspace{0.5in} $\displaystyle x$
\vspace{0.5in}
\caption{ Pion distribution amplitude for DLCQ cutoffs $K=10,12,15,20$, 
darker data points meaning larger $K$. $K \to \infty$ extrapolated curve
lies in the shaded region.  
\label{fig8}}
\end{figure}

\begin{figure}
\centering
$\displaystyle C_{\pi}$ \hspace{5pt}  \BoxedEPSF{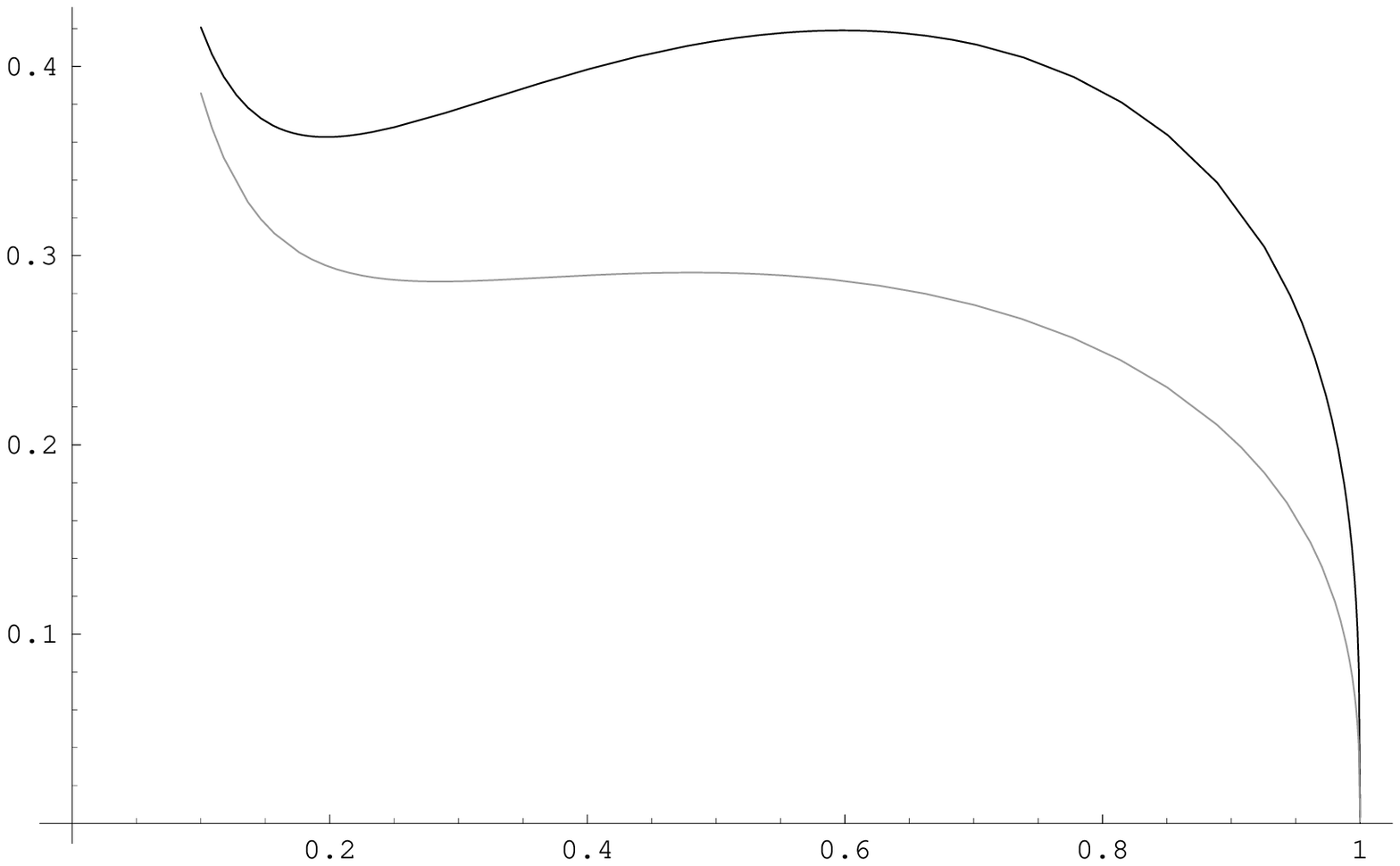 scaled 800}\\
\vspace{-1.7in} \hspace{0.5in} $\displaystyle x$
\vspace{0.5in}
\caption{Extrapolated quark helicity correlation function for the pion: black 
for anti-parallel helicities $C_{\pi}^{\rm anti}$; grey for parallel helicities
$C_{\pi}^{\rm para}$.
\label{fig9}}
\end{figure}

\end{document}